**Simplified formulas for the generation of terahertz waves from semiconductor surfaces excited with a femtosecond laser**

Masayoshi Tonouchi[a)], Institute of Laser Engineering, Osaka University, 2–6 Yamada-Oka, Suita-city, Osaka 565-0871 Japan

[a)] Author to whom correspondence should be addressed: tonouchi@ile.osaka-u.ac.jp

**ABSTRACT**

We derive simple formulas to explain terahertz (THz) emission from semiconductor surfaces excited by a femtosecond (fs) laser. Femtosecond optical pulses with energies larger than the bandgap create photocarriers which travel and generate THz radiation, according to the time derivative of the photocurrent. By assuming that only electrons traveling in an ultrafast time scale, less than a few hundred fs, contribute to THz radiation, one can obtain simple expressions for the emission originating from the photocarrier drift accelerated with a built-in field or from the photocarrier diffusion. The emission amplitude of the former is in proportion with electron mobility, the Schottky-Barrier height, and the laser intensity and one of the latter with the laser intensity and diffusion coefficient squared. We also discuss the formula for emission from metal-insulator-semiconductor structures. The derived expressions are useful in understanding the THz emission properties observed by a laser THz emission microscope (LTEM), bringing LTEM into real applications in the field of semiconductor research and development.

**I. INTRODUCTION**

Terahertz (THz) emission spectroscopy (TES) is an emerging tool that captures the ultrafast optical response of electronic materials upon illumination with a femtosecond (fs) laser.[1–3] Femtosecond optical pulses generate photocarriers in materials when pulse energy is larger than the energy bandgap, and ultrafast displacement of the photocarriers via their acceleration and/or diffusion induces THz wave radiation in accordance with the time derivative of the photocurrent. TES has been intensively studied for about three decades, initially for semiconductors,[4–6] and then for other kinds of materials.[7–10] Leitenstofer et al. has proven that TES, applying for GaAs and InP, provides complimentary and unique probes on the ultrafast carrier dynamics[11-13] to the conventional ultrafast characterization methods for the semiconductors such as photoluminescence (PL), pump-and-probe optical reflection measurements, and four-wave-mixing.[14] Although TES has been applied for the many semiconductors, there remain broad unexplored areas of research in the field of real devices, wide bandgap semiconductors, and so on.

Recently we have applied the TES to the various semiconductors. Examples are the evaluation of the surface potential of the passivated Si wafers,[15,16] which would bring the TES into the real semiconductor industrial applications, the study on the conduction band bending of the β-$Ga_2O_3$ near the surface with TES for the first time,[17] and the imaging of the spontaneous polarization in GaN.[18] The ultrafast optical response in the wide bandgap semiconductors is still an unexplored area where it is worth applying TES. However, the emission mechanism includes complex carrier dynamics[19-22], such as anisotropic optical responses, because of which we might have to evaluate them for each case by means of simulations such as Monte-Carlo method. For the semiconductor industrial applications, it is desirable to have a simplified modeling and capture instinct carrier dynamics.

In the present work, we derive simplified expressions for THz radiation from semiconductor surfaces, employing a short-time approximation. We are interested in the initial dynamic response just after fs-laser illumination, over the order of a few hundred fs, and neglect hole contributions due to their slow nature. We separately address the emission from surfaces based on photocarrier acceleration by the surface built-in fields and photocarrier diffusion. We also study the emission from metal-insulator-semiconductor (MIS) structures. We believe the derived formulas enable a better understanding of THz emission properties, especially in the wide-gap semiconductors and MIS devices.

## II. MODELS AND DISCUSSION

There are two major mechanisms of THz wave generation on semiconductor surfaces by means of ultrafast photocurrent generation with fs laser illumination. One is photocurrent generation due to photocarriers accelerated by a built-in field near the surface called a "drift current"; the other is photocarrier diffusion from the surface into the semiconductor inwards. In the latter case, differences in the mobility of electrons and holes induce a transient current called the "Photo-Dember effect." The drift photocurrent explains THz emission in normal semiconductors, such as Si, GaAs, and InP, whereas the diffusion one does the emission from narrow gap semiconductors, such as InAs, and InSb. Here we discuss simplified models for the above mechanisms. Typically we use fs lasers with a pulse width of 100 fs or less. Thus, for simplicity, we assume that the time range of interest is about a few hundred fs.

A. THz radiation by surge drift current

Figure 1(a) is a schematic band diagram of a semiconductor surface. The primary mechanism for THz generation due to surge drift current depends on the relationship between surface potential, $\psi_{BS}$, depletion layer thickness, $w$, diffusion voltage, $V_D$, mobility $\mu$, and optical penetration depth, $\lambda_L$. The time derivative of the photocurrent is a source for THz radiation. In the present case, one can neglect the initial carrier number and velocity of the electrons before laser excitation. Femtosecond illumination generates a photocurrent, $\Delta J$, which is proportional to the laser intensity, $I_p$, during the fast transient time scale, $\Delta t$, which should be around 200–300 fs. Then we can derive the simple formula:

$$E_{THz} \propto \frac{\partial J}{\partial t} \propto \frac{\Delta n \Delta v}{\Delta t} \propto \mu E_B I_p, \qquad (1)$$

where $E_B$ is the built-in field.

For $w > \lambda_L$、$E_B$ is approximately replaced by $E_{MAX}$, which is the built-in field at the surface. Then we obtain:

$$E_{THz} \propto \pm \mu \sqrt{\frac{N_i V_D}{\varepsilon_r}} I_p, \qquad (2)$$

where $N_i$ and $\varepsilon_r$ are the impurity density and dielectric constant of the semiconductors of interest, respectively. The sign of $E_{THz}$ changes according to carrier type, i.e., n-type and p-type, for the doped semiconductors,[23] or according to the band-bending direction near the surface for the nondoped/semi-insulating semiconductors. For $w \leq \lambda_L$, a rough approximation allows us to replace $I_p$ by $I_p \frac{w}{\lambda_L}$. Then, we obtain:

$$E_{THz} \propto \pm \mu \frac{V_D}{\lambda_L} I_p. \qquad (3)$$

which is the case for most doped semiconductors.

Figure 2 shows an example of THz emission waveforms from n-type and p-type InP wafers with carrier densities of around $5 \times 10^{18}$ cm$^{-3}$ and $1 \times 10^{18}$ cm$^{-3}$, respectively.[24] The signs of the waveforms differ depending on carrier type. For doped semiconductors, $V_D$ can be replaced by the surface potential (Schottky-Barrier (SB)), which is often measured using a Kelvin Force Microscope (KFM). In the case of the THz emission from InP given in Fig.2, the SB height and $w$ of p-InP are slightly larger than those of n-InP, but of the same order. Thus, the amplitude difference can also be presumably attributed to differences in electron mobility.

B. THz radiation by surge diffusion current

The THz emission originating in the diffusion of the photocarriers, i.e., the photo-Dember effect, has been studied for a long time. Many cases are the fs laser illumination on the narrow bandgap semiconductors.[20,23, 25-27] The situation realizes the hot photocarrier injection with the energy far above the conduction bands. Thus the THz emission is explained by the quasi-ballistic high-energy-carrier transport.

This case allows us to assume simply that in Eq. (1), $\Delta n$ and $\Delta v$ are replaced by $I_p$ and the acquired velocity from the excess photon energy, respectively. Thus we obtain,

$$E_{THz} \propto I_p \sqrt{\frac{E_p - E_g}{m^*}}. \qquad (4)$$

This formula indicates that the THz amplitude is a function of the laser intensity and square root of the excess energies divided by the effective mass $m^*$, but the sign keeps the same polarity for both the n-type and p-type semiconductors.[23,26]

We examined the THz emission from the narrow bandgap semiconductors, InGaAs (In: 53%), InGaAs (In: 60%), n-InAs, and InSb wafers, previously. Figure 3 gives an example of the THz emission power replotted from Ref. (26), which corresponds to the $E_{THz}$ doubled as a function of the excess energy measured at an excitation wavelength of 1560 nm and a power of 2.5 mW. Although they should have different electron masses, which would affect those electron velocities, the trend that the integrated amplitude increases linearly with the excess energy agrees with the Eq. (4). Although, of course, the emission has the strong anisotropic properties depending on the effective mass, bandstructure, scattering time, etc. of the materials,[20] it is almost predictable that the THz emission due to the diffusion increases notably with increasing excess energy.

Another case discussed here is that the photocarriers are excited with energies slightly above the bandgaps. The photocurrent induced by the electron density gradient can be a source of the electromagnetic wave generation. Since this proceeds with the thermalization, it would be a slow effect. However, in the polar semiconductors, the optical phonon scattering induces the quasi-thermalization proceeds in a few hundred femtoseconds, e.g., 240 fs in GaAs,[28] which is in the order of the present interest. Apostolopoulos *et al.* discussed the lateral photo-Dember effect in GaAs metal contact by the drift-diffusion equation in detail.[29] Heyman *et al.* reported that exciting GaAs with an energy higher than the energy gap shifts the origin of the emission mechanism from the drift current to diffusion.[30] However, the co-existence of the drift and diffusion current near the surface makes it difficult to study the mechanisms separately. As far as we know, there are no such studies to discuss the THz generation due to the pure carrier diffusion excited at the small excess energy. Thus it is worth describing the simple diffusion model here. Note that as discussed in the latter section, the THz emission from the flat band conditions can be studied in the MIS structures.

The diffusion current, $J_d$, is defined by $J_d = eD_n \frac{\partial n(x)}{\partial x}$, where $D_n$ is the electron diffusion coefficient, and, neglecting the drift current induced by the built-in field, the electron continuity equation is $\frac{\partial n(x)}{\partial t} = D_n \frac{\partial^2 n(x)}{\partial x^2} + G_n - \frac{(n(x)-n_0)}{\tau_n}$, where $G_n$ is a carrier generation rate, $\tau_n$ is the carrier lifetime, and $n_0$ is the initial electron density. For the present case for the THz excitation, one can assume that $\tau_n \approx \infty$ and $n_0 = 0$. Furthermore, it can be assumed that $G_n = I_0 exp(\frac{-x}{\lambda_L})$ at $t = 0$, and therefore, we can set the initial condition $n(x, t = 0) = I_0 exp(\frac{-x}{\lambda_L})$, where $I_0$ is the photon density per unit area. Accordingly, we can derive:

$$\frac{\partial n(x)}{\partial t} \approx D_n \frac{\partial^2 n}{\partial x^2} + I_0 exp(\frac{-x}{\lambda_L}) \approx D_n \frac{\partial^2 n}{\partial x^2}. \tag{5}$$

Inserting this into the diffusion current formula, we can obtain the THz E-field at $x$ as:

$$E_{THz}(x) \propto \frac{\partial J_d}{\partial t} = eD_n \frac{\partial^2 n}{\partial x \partial t} \sim eD_n^2 \frac{\partial^3 n}{\partial x^3} \tag{6}$$

Then, integrating $x$ from 0 to $\lambda_L$ using the short time approximation, the THz field is given by:

$$E_{THz} \propto D_n^2 \int_0^{\lambda_L} \frac{\partial^3 n}{\partial x^3} dx = \frac{D_n^2}{\lambda_L^2} I_0. \tag{7}$$

Since the total photo density is given by $I_p = 0.63 I_0 \lambda_L$, finally, we obtain the rough approximation formula as:

$$E_{THz} \propto \frac{D_n^2}{\lambda_L^3} I_p. \tag{8}$$

This formula indicates that the THz amplitude is a function of the diffusion coefficient squared and the laser intensity.

Since the diffusion coefficient is proportional to the mobility multiplied by "quasi-thermalized" electron temperatures, it is expected that the excess energy affects the emission properties. The electron temperature, $T_n$, increases with increasing photon energy, $E_p$, according to the equation:

$$\frac{3}{2} k_B T_n = E_p - E_g. \tag{9}$$

Neglecting the thermal energy at room temperature, 17meV, we obtain:

$$E_{THz} \propto \frac{\mu^2 T_n^2}{\lambda_L^3 I_p} \propto \mu^2 \frac{(E_p - E_g)^2}{\lambda_L^3} I_p. \tag{10}$$

One can assume that, near the conduction band edges, the mobility is constant or proportional to the square root of temperatures for the polar semiconductors.[31] Thus Eq. (10) suggests that the THz intensity increases as the excess energies doubled or cubed, respectively, which has a different relationship from Eq.(4). On this point, we need the experimental discussion by the wavelength dependence of the THz emission at the flat-band condition of the MIS structure.

In summary, the amplitude of the THz emission arising from the surge current increases proportionally with laser intensity—examples are given in Fig. 4.[24,26] At low laser fluences, the linear relationships between the intensities and laser power agree with Eqs. (3), (4), and (10).

C. THz emission from metal-insulator-semiconductor structure

The surface potential of semiconductors covered with insulators, such as AlN and $SiO_2$,

changes with surface defects, defects in the insulators, and the electron affinity of the insulators. Since the built-in field controls THz emission properties, it is important to formulate the THz radiation electric field using the built-in field. We can find an expression for the surface field in the MIS structure in many text books, such as Sze's semiconductor bible.[31] Figure 1(b) gives the band-diagram of an MIS structure. The near-surface potential is given by:

$$E(\psi_S) = \pm \frac{2k_B T}{e} \frac{1}{\sqrt{2}L_D} F(\psi_S) = \pm \frac{\sqrt{2}k_B T}{eL_D} F(\psi_S) \qquad (10)$$

with definitions of $\beta = \frac{e}{k_B T}$,

$$F(\psi_S) = \sqrt{[exp(-\beta\psi_S) - 1 + \beta\psi_S] + \frac{n_{p0}}{p_{p0}}[exp(\beta\psi_S) - 1 - \beta\psi_S]},$$

and $L_D = \sqrt{\frac{\varepsilon_s \varepsilon_0}{e\beta p_{p0}}}$.

$L_D$ is the extrinsic Debye length.[31] Thus, the THz radiation field is expressed by:

$$E_{THz} \propto \pm \frac{\sqrt{2}k_B T}{eL_D} F(\psi_{BS}) \qquad (11)$$

where $\psi_{BS}$ is the surface potential at the interface. Tending to the flat band condition, we obtain:

$$E_{THz} \propto \pm \frac{\sqrt{2}k_B T}{eL_D} \sqrt{[exp(-\beta\psi_S) - 1 + \beta\psi_S]}. \qquad (12)$$

Figure 5 shows the relationship between $E_{THz}$ and $\psi_S$ near the flat-band condition. The result explains the tendency of the observed relation in Refs. 24 and 26, except for the shift due to THz emission by diffusion under flat-band conditions expressed by Eq. (8).

For both positive bias and negative bias conditions, $E_{THz}$ saturates at certain values. With increasing bias, the inversion layer is formed, and $\psi_S$ reaches a maximum value close to $E_g$. One can replace $V_D$ by $E_g$ in Eqs. (2) and (3), depending on the relationship between $w$ and $\lambda_L$. With increasing negative bias, an accumulation layer is formed, and $\psi_S$ saturates at around $E_f - E_v$ for p-type semiconductors. Although the value of $E_f - E_v$ is small, $w$ is also small, resulting in a high built-in field near the interface of the insulator and semiconductor. We will discuss this case in the near future.

## III. CONCLUSIONS

We propose some simple formulas to explain the properties of THz radiation arising from semiconductor surfaces and MIS structures. THz emission properties elucidate important photoexcited carrier dynamics near semiconductor surfaces, and give insight into the near-surface conditions of semiconductors. These formulas make the LTEM powerful for real applications in the field of semiconductor R&D, particularly as a member of the noncontact, nondestructive evaluation family that includes PL and Raman spectroscopy. LTEM will be especially important for wide-gap semiconductors, which have various imperfections in themselves, in their interfaces, and in their passivation layers.

**ACKNOWLEDGMENTS**

The author acknowledges Drs. F. R. Bagsican and M. Suzuki for their technical assistance. This work is partially supported by JSPS KAKENHI, Grant Nos. JP18KK0140, JP18K18861, and JSPS Core-to-Core Program.

**DATA AVAILABILITY STATEMENT**

The data that support the findings of this study are available from the corresponding author upon reasonable request.

Figure Captions

FIG. 1. Schematics and parameter definitions of a typical band diagram for (a) n-type semiconductor and (b) metal/insulator/p-type semiconductor surfaces. Here we neglect defects inside the semiconductors and the insulator.

FIG. 2. THz radiation waveforms in the time domain emitted from p-type and n-type semiconductors. Amplitude and origin of time are assigned arbitrarily.

FIG. 3. Excess photon energy dependence of the THz power. The power was roughly estimated by the multiplication by the peak THz amplitude and the first THz pulse width.

FIG. 4. Laser power dependence of THz emission peak amplitude for various semiconductors excited at wavelengths of (a) 780 nm and (b) 1560 nm. The signs of amplitude in (a) indicate that the electrons excited in SI-GaAs and unintentionally doped n-InAs travel inward, whereas those in SI-InP travel outward to the surface, regardless of their THz emission mechanism.

FIG. 5. Calculated THz emission amplitude from MIS as a function of the surface potential near the flat band conditions. The trend coincides with the data in Ref. 23, except for the shift due to the carrier diffusion component.

Fig. 1

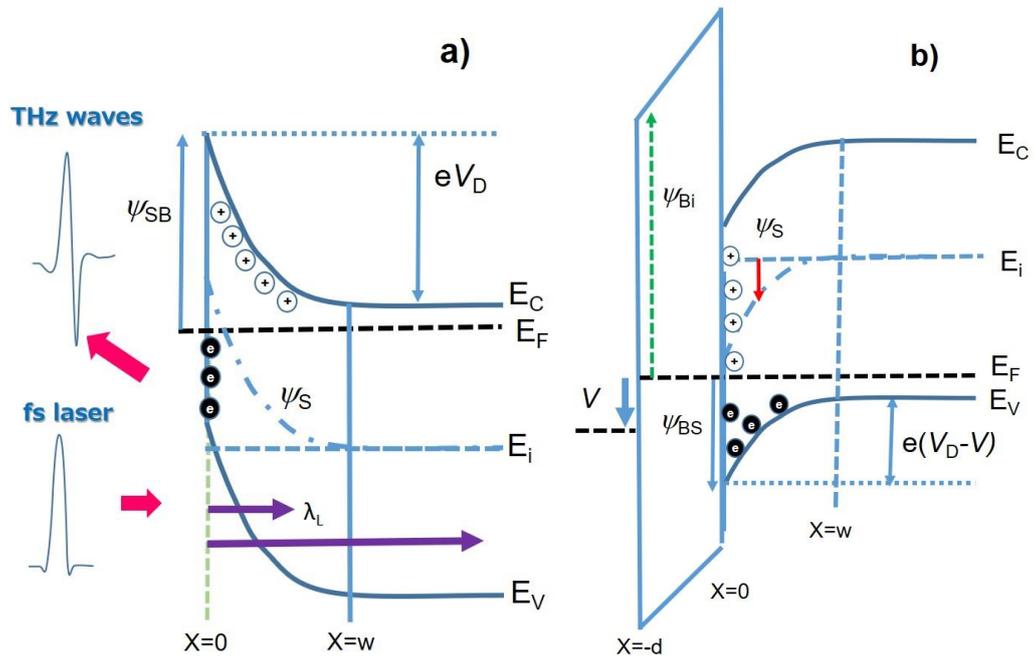

Fig. 2

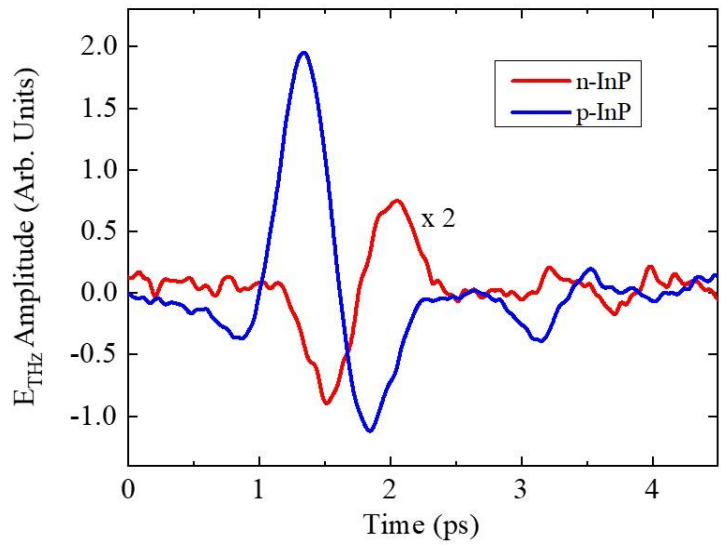

Fig.3

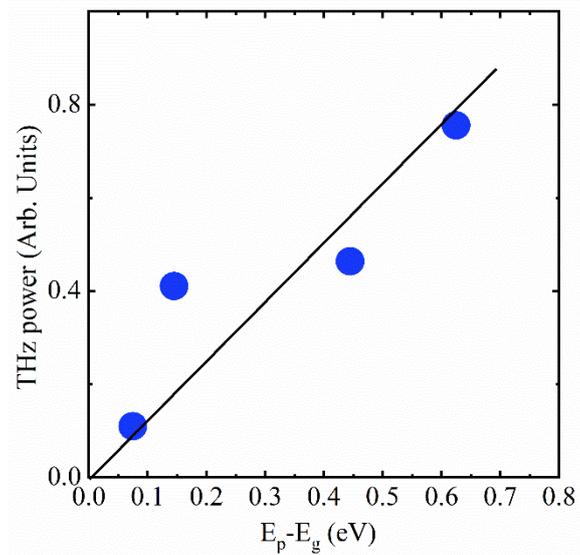

Fig. 4

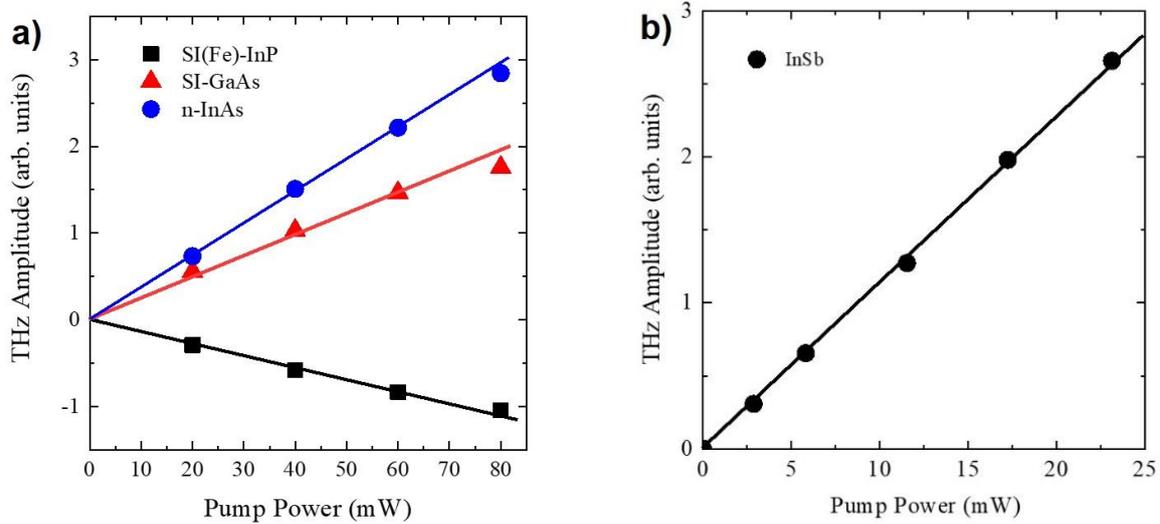

Fig. 5

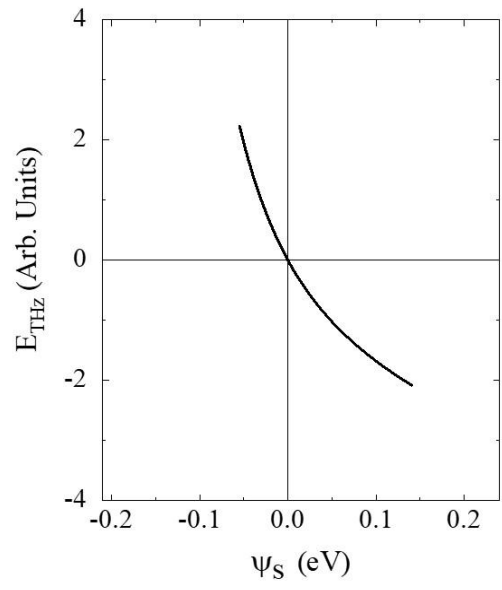